\documentclass[conference]{IEEEtran}
\IEEEoverridecommandlockouts
\usepackage{cite}
\usepackage[tbtags]{amsmath}
\usepackage{amssymb,amsfonts}
\usepackage{algorithmic}
\usepackage{graphicx}
\usepackage{textcomp}
\usepackage{xcolor}
\usepackage{caption}
\usepackage{commath}

\def\BibTeX{{\rm B\kern-.05em{\sc i\kern-.025em b}\kern-.08em
    T\kern-.1667em\lower.7ex\hbox{E}\kern-.125emX}}

\newtheorem{theorem}{\textbf{Theorem}}

\begin{document}

\title{Unknown Interference Modeling for Rate Adaptation in Cell-Free Massive MIMO Networks}
\author{\thanks{This work was supported by the FFL18-0277 grant from the Swedish Foundation for Strategic Research.}
\IEEEauthorblockA{Mahmoud Zaher\IEEEauthorrefmark{1}, Emil Björnson\IEEEauthorrefmark{1}, and Marina Petrova\IEEEauthorrefmark{1}\IEEEauthorrefmark{2}\\
 \{mahmoudz, emilbjo, petrovam\}@kth.se\\}

\IEEEauthorblockA{
\IEEEauthorrefmark{1}Department of Computer Science, KTH Royal Institute of Technology, Sweden\\
\IEEEauthorrefmark{2}Mobile Communications and Computing, RWTH Aachen University, Germany
}
}
\maketitle

\begin{abstract}

Co-channel interference poses a challenge in any wireless communication network where the time-frequency resources are reused over different geographical areas. The interference is particularly diverse in cell-free massive multiple-input multiple-output (MIMO) networks, where a large number of user equipments (UEs) are multiplexed by a multitude of access points (APs) on the same time-frequency resources. For realistic and scalable network operation, only the interference from UEs belonging to the same serving cluster of APs can be estimated in real-time and suppressed by precoding/combining. As a result, the unknown interference arising from scheduling variations in neighboring clusters makes the rate adaptation hard and can lead to outages. This paper aims to model the unknown interference power in the uplink of a cell-free massive MIMO network. The results show that the proposed method effectively describes the distribution of the unknown interference power and provides a tool for rate adaptation with guaranteed target outage.
\end{abstract}


\vspace{-2pt}\section{Introduction}

\vspace{-1pt}Wireless communication is continuously evolving to cope with the increasing traffic demands in mobile networks. Recently, three leading concepts have been intensively investigated to boost the capacity of mobile networks: achieving higher spectral efficiency (SE), allocating more bandwidth, and network densification \cite{zaher2022soft}. Massive multiple-input multiple-output (MIMO) offers substantial SE gains through spatial multiplexing of a large number of user equipments (UEs) on the same time-frequency resources \cite{zhang2019learning}. Cell-free massive MIMO represents a combination of cooperative principles with massive MIMO processing to allow for more network densification. It poses as a potential candidate for beyond 5G mobile communications due to its ability to utilize macro diversity and suppress multi-user interference to provide uniform service to the UEs \cite{demir2021foundations}. A cell-free network comprises a large number of distributed access points (APs) jointly serving the UEs in a given coverage area with no cell boundaries \cite{zaher2022soft,demir2021foundations}. For scalable network operation, each UE is only served by a subset of the APs in the network \cite{interdonato2019scalability}. The APs can either be divided into pre-defined AP clusters that the UE can choose from, or into user-centric clusters that are dynamically adapted to the current channel gains between the UE and all its surrounding APs. In both cases, though the combination of cell-free massive MIMO and network densification can offer high data rates, the strength of the co-channel interference from neighboring AP clusters can vary rapidly due to scheduling decisions and cause outages \cite{zaher2023bayesian}. Massive signaling is required to measure the instantaneous interference level; thus, it is usually not done in practice \cite{zhang2019learning}.

Prior works have focused on determining the distribution of the signal-to-interference-plus-noise ratio (SINR) in different MIMO setups. For the cellular MIMO setup, the original work \cite{gao1998theoretical} derives the cumulative distribution function (CDF) of the SINR with equal power interferers, whereas the SINR distribution in a Poisson field of interferers is derived in \cite{ali2010performance}, both assuming minimum mean square error (MMSE) combining and perfect CSI. The recent work \cite{zhang2019learning} derives an asymptotic approximation of the SINR distribution in a Poisson field of interferers with imperfect CSI by fitting a Gamma distribution to the co-channel interference. In addition, \cite{bai2016analyzing} provides an asymptotic approximation for the SINR distribution using an exclusion ball model with maximum ratio (MR) and zero-forcing (ZF) receivers. However, the method represents a good approximation only when the number of antennas is much greater than the number of interfering UEs. The work in \cite{lim2019distribution} utilizes regular patterns to derive the SINR distribution under uncorrelated Rayleigh fading, equal power interferers, and perfect CSI for an MMSE MIMO system. In addition, \cite{li2005distribution,ma2008capacity} approximate the SINR distribution by a Gamma random variable when using ZF and MMSE receivers with perfect CSI for the equal power case and uniform interferers, respectively. Similarly, in the cell-free setup, the initial works \cite{tan2022performance,kurma2022outage} fit a Gamma distribution for each of the desired signal and interference to approximate the resulting SINR distribution.
The common approach in these works is either knowledge of the statistical properties of the underlying interference for computing the outage probability, which leads to expressions that are only valid for a particular set of assumptions; for example, precoding/combining and interference channel distribution; or fitting a single Gamma approximation to the SINR or interference which lacks a clear theoretical motivation and does not necessarily work well for all network configurations.

In this paper, we extend on our primal work \cite{zaher2023bayesian}, which utilizes a Bayesian method to infer the distribution of the unknown interference power in a single-cell setup, to derive the distribution of the total unknown interference power arising from UEs in neighboring clusters in the cell-free massive MIMO setup. The current work addresses a more challenging scenario where the combined unknown interference from the APs perceived at the central processing unit (CPU) negatively affects the SINR. We utilize observations of the unknown interference power at each AP in the serving cluster separately in a relatively large network to compute the distribution parameters for the unknown interference power at each AP. Moreover, we derive an expression for the CDF of the total unknown interference power based on the characteristic functions of the unknown interference power at each AP. We verify the tightness of the semi-analytical expressions numerically and provide a tool for UL rate adaptation with guaranteed target outage performance based on our analytical work.

\section{System Model}\label{model}

The considered cell-free massive MIMO network consists of $L$ APs and $K$ single-antenna UEs. Each AP is equipped with an array of $N$ antennas. The UEs are divided into two different sets: the set of known UEs $\mathcal{D}_n$ containing the desired UE and $K_n$ known interferers belonging to the same serving AP cluster such that $|\mathcal{D}_n| = K_n +1$, and the set of unknown interferers $\mathcal{D}_u$ with $|\mathcal{D}_u| = K_u$. Accordingly, we have the total number of UEs in the system as $K = K_n + K_u + 1$. Further, we assume that the serving AP cluster for the desired UE and known interferers comprises $L_s$ APs from the total set of APs in the network. For simplicity and without loss of generality, we assume non-overlapping AP clusters meaning that a UE is either a known or unknown interferer to the $L_s$ APs altogether. Following the standard block-fading channel model, the time-varying wideband channels are divided into time-frequency coherence blocks with static and frequency-flat channels. Each coherence block comprises $\tau_c$ symbols and the channels take independent random realizations in each block. The channel between UE $k$ and AP $l$ is represented by $\mathbf{h}_{kl} \in \mathbb{C}^{N \times 1}$ and modelled by correlated Rayleigh fading as $\mathbf{h}_{kl} \sim \mathcal{N}_{\mathbb{C}}(\mathbf{0}, \mathbf{R}_{kl})$, where $\mathbf{R}_{kl} \in \mathbb{C}^{N \times N}$ represents the spatial correlation matrix and $\beta_{kl} = \frac{1}{N} \hspace{1pt}\textrm{tr}\!\left(\mathbf{R}_{kl}\right)$ provides the average channel gain.

\subsection{Channel Estimation}

We consider a pilot transmission phase for channel estimation followed by a data transmission phase. To that end, the coherence block is divided into $\tau_p$ symbols for UL pilots and $\tau_u$ symbols for UL data transmission, i.e., $\tau_c = \tau_p + \tau_u$.

Each UE is assigned a $\tau_p$-length pilot from a set of $\tau_p$ mutually orthogonal pilots utilized by the APs for channel estimation. We consider the general case where $\tau_p < |\mathcal{D}_n|$ such that pilot contamination exists among UEs belonging to the same serving AP cluster, and adopt a simple pilot assignment algorithm from \cite{demir2021foundations} that effectively avoids the worst-case situations. Since this is not the main focus of the paper, the details of the algorithm are omitted due to space limitations. We let $\mathcal{P}_t \subset \{1, \hdots, K\}$ denote the UEs that are assigned to pilot $t$. For estimation of the UEs' channels belonging to $\mathcal{P}_t$, we correlate the received signal at AP $l$ with the pilot $t$. As a result, the signal $\mathbf{y}_{tl}^p \in \mathbb{C}^{N \times 1}$ becomes
\begin{equation}
\mathbf{y}_{tl}^p = \sum_{i \in \mathcal{P}_t}\sqrt{\tau_pp_i}\mathbf{h}_{il} + \mathbf{n}_{tl}
\end{equation}
where $p_i\hspace{-1pt}$ denotes the transmit power of UE $i$ and $\mathbf{n}_{tl} \hspace{-0.85pt}\sim \mathcal{N}_{\mathbb{C}}\hspace{-0.5pt}\left(\mathbf{0}, \sigma^2\mathbf{I}_N\hspace{-0.5pt}\right)$ is the additive Gaussian noise vector at AP $l$.

We assume that only the channel statistics of the UEs' belonging to $\mathcal{D}_n$ are known at the $L_s$ APs serving the desired UE. To that end, we employ an estimator that utilizes only the known channel statistics and has a similar structure to the MMSE estimator in \cite{demir2021foundations}. Hence, the channel between a known UE $k$, $k \in \mathcal{D}_n$ and AP $l$, $l = 1, \hdots, L_s$ is estimated as

\begin{equation} \label{eq:channel-estimate}
\hat{\mathbf{h}}_{kl} = \sqrt{\tau_pp_k}\mathbf{R}_{kl}\left(\sum_{i \in \mathcal{P}_t \cap \mathcal{D}_n}\hspace{-5pt}\tau_pp_i\mathbf{R}_{il} + \sigma^2\mathbf{I}_N\right)^{\hspace{-1pt}-1}\hspace{-1pt}\mathbf{y}_{tl}^p,
\end{equation}
where $\cap$ denotes the intersection of sets.
We highlight that the expression in the inverse in \eqref{eq:channel-estimate} corresponds to the covariance matrix of the received pilot signal in the absence of unknown interferers; that is, the above estimate is MMSE optimal in the absence of unknown interferers. Note that AP $l$, $l = 1, \hdots, L_s$ is only capable of utilizing the channel statistics of UE $k$, $k \in \mathcal{D}_n$ for determining the estimates, as the channel statistics of the unknown interferers that may share the same pilot as UE $k$ are not available at the APs. 

\subsection{Uplink Data Transmission}

Generally, in a cell-free massive MIMO system, all UEs transmit their signals on the same time-frequency resources. Hence, the received UL signal at AP $l$ is given by
\begin{equation}
    \mathbf{y}_l^{ul} = \sum_{i = 1}^{K}\sqrt{p_{i}}\mathbf{h}_{il}s_i + \mathbf{n}_l,
\end{equation}
where $s_i$ denotes the zero-mean unit-power signal of UE $i$ and $\mathbf{n}_l \sim \mathcal{N}_{\mathbb{C}}\!\left(\mathbf{0}, \sigma^2\mathbf{I}_N\right)$ represents the additive noise at AP $l$. Each AP locally processes its received signal by computing local estimates of the data symbols which are then passed to the CPU of the cluster for final decoding. A linear receive combining scheme is adopted by each AP in the serving cluster of the desired UE $k$; thus, the local signal estimate at AP $l$ for the desired UE $k$ is calculated by
\begin{equation}
    \hat{s}_{kl} = \sqrt{p_k}\mathbf{v}_{kl}^H\mathbf{h}_{kl}s_k + \sum_{\substack{i = 1\\i \neq k}}^{K}\sqrt{p_i}\mathbf{v}_{kl}^H\mathbf{h}_{il}s_i + \tilde{n}_{kl},
\end{equation}
where $\mathbf{v}_{kl}$ corresponds to the receive combining vector by AP~$l$ that is designated for UE $k$ and $\tilde{n}_{kl} = \mathbf{v}_{kl}^H\mathbf{n}_l$ represents the processed noise. It is to be noted that AP $l$ can only utilize the locally estimated channels of the known interferers $\{\hat{\mathbf{h}}_{il}: i \in \mathcal{D}_n\}$ in its design of $\mathbf{v}_{kl}$. In this paper, we employ the MR and local partial RZF as the receive combining vectors for the numerical evaluation, however, we stress that the system model is applicable with any other combining scheme. As such, the receive combining vectors are computed as
\begin{equation}
\mathbf{v}_{kl} =
\begin{cases}
      \hat{\mathbf{h}}_{kl}/\|\hat{\mathbf{h}}_{kl}\|^2 & \textrm{for MR}, \\
      \left(\sum\limits_{i \in \mathcal{D}_n}p_i\hat{\mathbf{h}}_{il}\hat{\mathbf{h}}_{il}^H + \sigma^2\mathbf{I}_N\right)^{-1}p_k\hat{\mathbf{h}}_{kl} & \textrm{for RZF}.
\end{cases}
\end{equation}

Afterwards, the CPU receives the local signal estimates of the desired UE $k$ from the serving AP cluster $\{\hat{s}_{kl}: l = 1, \hdots, L_s\}$. The CPU of the cluster then performs linear combining of the local signal estimates using the weights $\{a_{kl}: l = 1, \hdots, L_s\}$ to compute the signal estimate $\hat{s}_k$ of the desired UE $k$ as
\begin{align}
    \hat{s}_k &= \sum_{l = 1}^{L_s}a_{kl}^*\hat{s}_{kl}\notag\\
    &= \sum_{l = 1}^{L_s}a_{kl}^*\sqrt{p_k}\mathbf{v}_{kl}^H\mathbf{h}_{kl}s_k + \sum_{l = 1}^{L_s}a_{kl}^*\sum_{\substack{i = 1 \\ i \neq k}}^K\sqrt{p_i}\mathbf{v}_{kl}^H\mathbf{h}_{il}s_i + \tilde{n}_k\notag\\
    &= \mathbf{a}_k^H\mathbf{g}_{kk}s_k + \mathbf{a}_k^H\sum_{\substack{i = 1 \\ i \neq k}}^K\mathbf{g}_{ki}s_i + \tilde{n}_k,
\end{align}
where $\mathbf{a}_k = \left[a_{k1}, \hdots, a_{kL_s}\right]^T \in \mathbb{C}^{L_s}$ is the weighting vector at the CPU, $\mathbf{g}_{ki} = \bigl[\sqrt{p_i}\mathbf{v}_{k1}^H\mathbf{h}_{i1} ,\hdots ,\sqrt{p_i}\mathbf{v}_{kL_s}^H\mathbf{h}_{iL_s}\bigr]^T \in \mathbb{C}^{L_s}$ represents the receive-combined channels at the serving AP cluster, and $\tilde{n}_k = \sum_{l = 1}^{L_s}a_{kl}^*\tilde{n}_{kl}$. Note that the weighting vector $\mathbf{a}_k$ can only utilize the known channel statistics at the CPU. By leveraging the large-scale fading decoding (LSFD) approach of the cell-free massive MIMO system as in \cite{demir2021foundations,bjornson2019making}, an achievable SE can be computed by characterizing the desired signal based on the deterministic non-zero average $\mathbf{a}_k^H\mathbb{E}\{\mathbf{g}_{kk}\}$, which can be assumed known at the CPU. Hence, the achievable UL SE of the desired UE $k$ under LSFD is 
\begin{equation} \label{eq:SEk}
\textrm{SE}_k = \frac{\tau_u}{\tau_c}\textrm{log}_2\left(1 + \textrm{SINR}_k\right),
\end{equation}
where
\begin{equation}
\textrm{SINR}_k = \frac{|\textrm{DS}_k|^2}{\textrm{IUI}_k^u + \textrm{IUSI}_k^n + \mathbf{a}_k^H\mathbf{F}_k\mathbf{a}_k}
\label{SINR}
\end{equation}
and
\begin{align}
    &\textrm{DS}_k = \mathbf{a}_k^H\mathbb{E}\{\mathbf{g}_{kk}\},\\
    &\textrm{IUI}_k^u = \sum_{i \in \mathcal{D}_u}\mathbb{E}\{|\mathbf{a}_k^H\mathbf{g}_{ki}|^2\},\label{unknownint}\\
    &\textrm{IUSI}_k^n = \sum_{i \in \mathcal{D}_n}\mathbb{E}\{|\mathbf{a}_k^H\mathbf{g}_{ki}|^2\} - |\mathbf{a}_k^H\mathbb{E}\{\mathbf{g}_{kk}\}|^2,\\
    &\mathbf{F}_k = \sigma^2\textrm{diag}\left(\mathbb{E}\{\|\mathbf{v}_{k1}\|^2, \hdots, \|\mathbf{v}_{kL_s}\|^2\}\right).
\end{align}

In the above, $\textrm{SINR}_k$ is the effective SINR, $\textrm{DS}_k$ represents the desired signal over the deterministic average channel, $\textrm{IUI}_k^u$ represents the unknown inter-user interference power arising from UEs in neighboring clusters, whereas $\textrm{IUSI}_k^n$ represents the inter-user interference power from the known interferers plus the self-interference due to channel uncertainty. $\mathbf{F}_k$ denotes the covariance matrix of the processed noise. The weight vector is selected in a similar manner to \cite{bjornson2019making}, however, utilizing only the known channel statistics at the CPU
\begin{equation}
    \mathbf{a}_k = \left(\sum_{i \in \mathcal{D}_n}\mathbb{E}\{\mathbf{g}_{ki}\mathbf{g}_{ki}^H\} + \mathbf{F}_k\right)^{-1}\mathbb{E}\{\mathbf{g}_{kk}\}.
\end{equation}

\section{Unknown Interference Power Distribution}

This work focuses on characterizing the distribution of the total unknown interference power that arises from UEs in neighboring clusters. In this section, we extend our idea for the single-cell setup in \cite{zaher2023bayesian} to the cell-free massive MIMO setup. In practice, the unknown interference power from adjacent AP clusters, i.e., UEs not served by the same AP cluster as the desired UE, is not known in the desired UE's serving cluster, since it can change rapidly due to user mobility and scheduling decisions in the neighboring clusters \cite{zhang2019learning,zaher2023bayesian}. In our previous work \cite{zaher2023bayesian}, we have derived the distribution for the unknown interference power at a single AP as an Inverse-Gamma distribution. In order to characterize the total unknown interference power $\textrm{IUI}_k^u$ term in \eqref{unknownint}, we expand the term as
\begin{equation}
    \textrm{IUI}_k^u = \sum_{i \in \mathcal{D}_u}\mathbb{E}\{|\mathbf{a}_k^H\mathbf{g}_{ki}|^2\} = \mathbf{a}_k^H\left(\sum_{i \in \mathcal{D}_u}\mathbb{E}\{\mathbf{g}_{ki}\mathbf{g}_{ki}^H\}\right)\mathbf{a}_k,
\end{equation}
The summation in the parentheses represents the covariance matrix of the unknown interference power at the $L_s$ serving APs for the desired UE $k$. The diagonal terms of this matrix are the unknown interference powers at each AP, whereas the off-diagonal elements represent the cross-covariance between the unknown interference at the different APs. A typical AP-UE association strategy, as shown for pre-defined AP clusters in Fig.~\ref{scenario}, results in the strongest contributors to the unknown interference being different for the different APs; that is, the largest terms in the sum $\sum_{i \in \mathcal{D}_u}\sqrt{p_i}\mathbf{v}_{kl}^H\mathbf{h}_{il}s_i$ will be different for the different APs due to the pathloss differences. This results in the cross-covariance of the unknown interference being negligible. In other words, the covariance matrix will be diagonal with the unknown interference power at each AP as the diagonal elements.
Accordingly, the total unknown interference power can be approximated by $\textrm{IUI}_k^u \approx \sum_{l = 1}^{L_s}|a_{kl}|^2\textrm{IUI}_{kl}^u$, where $\textrm{IUI}_{kl}^u$ corresponds to the unknown interference power at AP $l$. The total unknown interference power $\textrm{IUI}_k^u$ then comprises a weighted sum of the different unknown interference powers experienced at each AP of the serving cluster. Building on our previous result for the single-cell scenario in \cite{zaher2023bayesian}, then for a dense network, the component terms $\{\textrm{IUI}_{kl}^u: l = 1, \hdots, L_s\}$ can be modeled as independent non-identically distributed Inverse-Gamma random variables. We denote the shape and scale parameters for the $l^{\textrm{th}}$ component by $\alpha_l$ and $\beta_l$, respectively, and so we have $\textrm{IUI}_{kl}^u \sim \textrm{IG}\left(\alpha_l, \beta_l\right)$. We stress that this approach is particularly useful as different unknown interferers can be modeled separately at each serving AP, regardless of the utilized clustering method.

For computing the parameters of the Inverse-Gamma random variables, the statistics of the unknown co-channel interferers' power can be estimated separately by each AP in the serving cluster. Through observing a sufficiently large sample of the interference powers and with the viable assumption that the statistics of the unknown interferers' power vary slowly over the day compared to the desired UE's transmission interval, the sample mean and sample variance of the unknown interference power converge to the true mean and variance of the distribution, respectively \cite{zhang2019learning,zaher2023bayesian}.

In the following, we derive the CDF of the total unknown interference power at the CPU as a weighted sum of independent non-identically distributed Inverse-Gamma random variables.

\begin{theorem}
\textit{The CDF of the total unknown interference power is given by}
\begin{align}
    &F_{\textrm{IUI}_k^u}\left(x\right) \approx \frac{1}{2} - \frac{1}{\pi}\int_0^{\infty}\textrm{Im}\left(\frac{e^{-jtx}\phi\left(t\right)}{t}\right)dt,\label{inv_formula}\\
    &\phi\left(t\right) = \prod_{l = 1}^{L_s}\phi_l\left(|a_{kl}|^2t\right),\label{char_fun_tot}\\
    &\phi_l\left(t\right) = \frac{2\left(-j\beta_lt\right)^{\alpha_l/2}}{\Gamma\left(\alpha_l\right)}K_{\alpha_l}\left(2\left(-j\beta_lt\right)^{1/2}\right),\\
    &\alpha_l = \left(\frac{\mu_l^2}{v_l}\right) + 2,\qquad\qquad\beta_l = \left(\frac{\mu_l^2}{v_l} + 1\right) \mu_l,
    \label{dist_par}
\end{align}
\textit{where $\phi\left(t\right)$ represents the characteristic function of the total unknown interference power at the CPU, $\phi_l\left(t\right)$ denotes the characteristic function of the Inverse-Gamma random variable representing the unknown interference power at the $l^{th}$ AP, $K_{\alpha}\left(\cdot\right)$ denotes the modified Bessel function of the second kind, and $\mu_l$ and $v_l$ represent the estimated mean and variance for the unknown interference power at AP $l$.}
\begin{IEEEproof}
The proof is given in Appendix A.
\end{IEEEproof}
\end{theorem}

The version of the inversion formula by \cite{gil1951note} in \eqref{inv_formula} is particularly useful for the numerical evaluation of the CDF of the total unknown interference power via a one-dimensional numerical integration \cite{witkovsky2001computing}. Another advantage of this method is that the unknown interference power can be dealt with locally at each AP and only its statistics are required to be communicated from the APs to the CPU to compute the respective distribution parameters. In the numerical evaluation, we will show the tightness of the derived distribution for the total unknown interference power. We present the result for different numbers of unknown interferers and different desired UE locations.

\section{{\large$\epsilon$}-Outage Rate}

The main goal of this paper is to develop a method for robust rate adaptation that can combat rapid variations in the total unknown interference power at the CPU. The unknown interference raises an issue for robust rate adaptation as the achievable SE in \eqref{eq:SEk} cannot be computed. The reason is that the term $\mathrm{IUI}_k^u$ in \eqref{unknownint} relies on the average fading coefficients of UEs in neighboring AP clusters that are unknown to the serving cluster of the desired UE. The randomness in this term comes from the unknown UEs' locations and respective unknown channel statistics.
In practice, the instantaneous realizations of these interference terms are not measured at the serving AP cluster of the desired UE as it would require large signaling overhead \cite{zhang2019learning} and because scheduling decisions are made simultaneously in all cells. In the following, we exploit our derived analytical results to attain the outage probability and $\epsilon$-outage SE performance for the uplink of the cell-free massive MIMO system described in Section \ref{model}. We stress that the considered outage framework models the risk of selecting a rate/SE that is unsupported due to large unknown interference statistics, which is fundamentally different from traditional outage analysis that considers the scenario of having unknown fading realizations for the desired signal. We note that the traditional scenario is not applicable to massive MIMO networks due to the channel hardening phenomena \cite{demir2021foundations} that reduces the fluctuations in the desired signal gain, and the coding that spans over many fading realizations.

The CDF for the SINR expression in \eqref{SINR} is given by
\begin{equation}
\begin{split}
    \textrm{Pr}\left[\textrm{SINR}_k \leq T\right] &= \textrm{Pr}\left[\frac{|\textrm{DS}_k|^2}{\textrm{IUI}_k^u + \textrm{IUSI}_k^n + \mathbf{a}_k^H\mathbf{F}_k\mathbf{a}_k} \leq T\right]\\
    &= 1 - F_{\textrm{IUI}_k^u}\left(\frac{|\textrm{DS}_k|^2}{T} - \textrm{IUSI}_k^n - \mathbf{a}_k^H\mathbf{F}_k\mathbf{a}_k\right)\!,
\end{split}
\end{equation}
where $F_{\textrm{IUI}_k^u}\left(\cdot\right)$ denotes the CDF of the weighted sum of Inverse-Gamma random variables representing the total unknown interference power $\textrm{IUI}_k^u$. To that end, given a target outage probability $\epsilon$, the $\epsilon$-outage SE is computed as
\begin{equation}
    R_k\left(\epsilon\right) = \frac{\tau_u}{\tau_c}\textrm{log}_2\left(1 +T_k\left(\epsilon\right)\right),
    \label{outage_rate}
\end{equation}
where
\vspace{-0.5em}
\begin{equation}
    T_k\left(\epsilon\right) = \frac{|\textrm{DS}_k|^2}{F_{\textrm{IUI}_k^u}^{-1}\left(1 - \epsilon\right) + \textrm{IUSI}_k^n + \mathbf{a}_k^H\mathbf{F}_k\mathbf{a}_k}.
    \label{outage_SINR}
\end{equation}
All terms, other than the unknown interference power, are assumed to be quasi-static statistical parameters, having fixed known values at the CPU for the serving AP cluster of the desired UE in a given transmission interval.

The proposed method for robust rate adaptation can be summarized in the following four-step procedure.

\begin{itemize}    
    \item \textbf{Step\,\,1:} Measure the total unknown interference power $\textrm{IUI}_{kl}^u$ at each AP for a certain number of samples. This can be done at slots where the known UEs are silent.
    
    \item \textbf{Step\,\,2:} Compute $\mu_l$ and $v_l$ for $\textrm{IUI}_{kl}^u$ using the samples in Step 1 at AP $l$, $l = 1, \hdots, L_s$ and send to the CPU.
    
    \item \textbf{Step\,\,3:} Compute $\alpha_l$ and $\beta_l$ according to \eqref{dist_par}.
    
    \item \textbf{Step\,\,4:} Compute the achievable SE for a given target outage probability $R_k\left(\epsilon\right)$ according to \eqref{outage_rate} and \eqref{outage_SINR}.
\end{itemize}

\section{Numerical Evaluation}

\begin{figure}
\centering
\setlength{\abovecaptionskip}{0.2cm plus 0pt minus 0pt}
\vspace{-0.51em}
\includegraphics[scale=0.49]{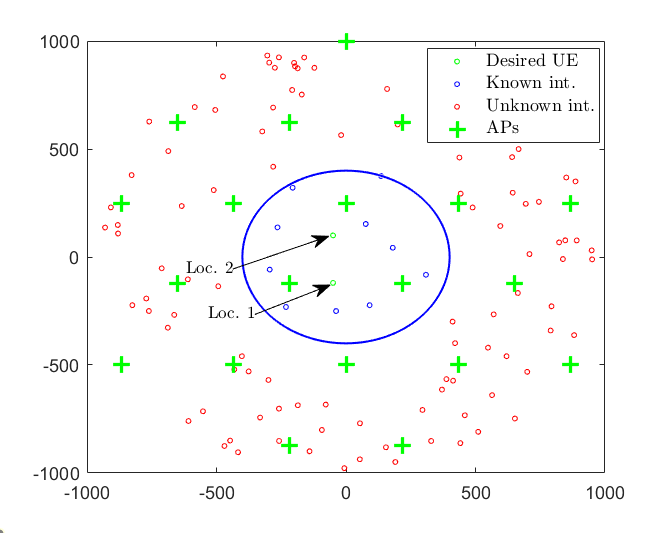}
\caption{Site map and UE distribution with $K_u = 100$ unknown interferers. Serving cluster is located inside the marked circle.}
\label{scenario}
\vspace{-1em}
\end{figure}

In this section, we validate the derived distribution for the total unknown interference power by Monte Carlo simulations and determine the achievable UL SE that satisfies different target outage probabilities. Each AP is equipped with a half-wavelength spaced uniform linear array of $N = 16$ antennas. We consider a serving AP cluster of $L_s = 3$ APs and the first tier of $6$ neighboring clusters comprising $18$ APs. Since the focus of the paper is on the SINR variations caused by the inter-cluster interference that is unknown at the serving AP cluster, the locations of the desired UE and known interferers are assumed fixed throughout the simulations. The simulation setup is depicted in Fig. \ref{scenario}. The figure shows a single random realization of the locations of the unknown interferers that are assumed to be associated with the neighboring AP clusters.
We employ two different simulations with the desired UE locations marked in Fig. \ref{scenario}. In both cases, $K_n = 10$ known interferers are at fixed locations that are chosen in a uniformly random manner within a circle of radius $400$\,m, where the serving AP cluster of the desired UE is the $L_s = 3$ APs inside the marked circle. In each simulation step, $K_u$ unknown interferers are randomly and uniformly dropped on a disk with the radius range of $[450, 1000]$\,m. The network simulation parameters are summarized in Table \ref{params}. We adopt the 3GPP Urban Microcell model for generating large-scale fading coefficients with correlated shadowing among the UEs, such that the large-scale fading coefficients are given by
\begin{equation}
\beta_{kl} = -30.5 - 36.7 \textrm{log}_{10}\left(\frac{d_{kl}}{1\,\textrm{m}}\right) + F_{kl} \hspace{2pt}\textrm{dB},
\label{pathloss}
\end{equation}
where $d_{kl}$ is the distance between UE $k$ and AP $l$ and $F_{kl} \sim \mathcal{N}\left(0, 4^2\right)$ represents the shadow fading. The shadowing is correlated between a single AP and different UEs as \cite{bjornson2019making}
\begin{equation}
\mathbb{E}\{F_{kl}F_{ij}\} = 
\begin{cases}
    4^22^{-\delta_{ki}/9\,\textrm{m}} & \textrm{for } l = j, \\
    0 & \textrm{for } l \neq j,
\end{cases}
\label{shadowing}
\end{equation}
where $\delta_{ki}$ denotes the distance between UE $k$ and UE $i$. Note that the correlation between the shadowing at different APs, which is the second case in \eqref{shadowing}, will be negligible because of the relatively larger inter-AP distances compared to the distances between the UEs.

\begin{table}
\begin{center}
\caption{Network simulation parameters.}
\begin{tabular}{ |c|c| }
\hline
Bandwidth & $20$\,MHz \\
Number of serving APs & $L_s = 3$ \\
Total number of APs & $L = 21$ \\
Number of antennas per AP & $N = 16$ \\
Number of known interferers & $K_n = 10$ \\
Pathloss exponent & $\alpha = 3.67$ \\
UL transmit power & $p_i = 100$\,mW \\
UL noise power & $-94$\,dBm \\
Coherence block length & $\tau_c = 200$ \\
Pilot sequence length & $\tau_p = 10$ \\
\hline
\end{tabular}
\label{params}
\end{center}
\vspace{-1em}
\end{table}

\begin{figure}
\centering
\setlength{\abovecaptionskip}{0.33cm plus 0pt minus 0pt}
\includegraphics[scale=0.47]{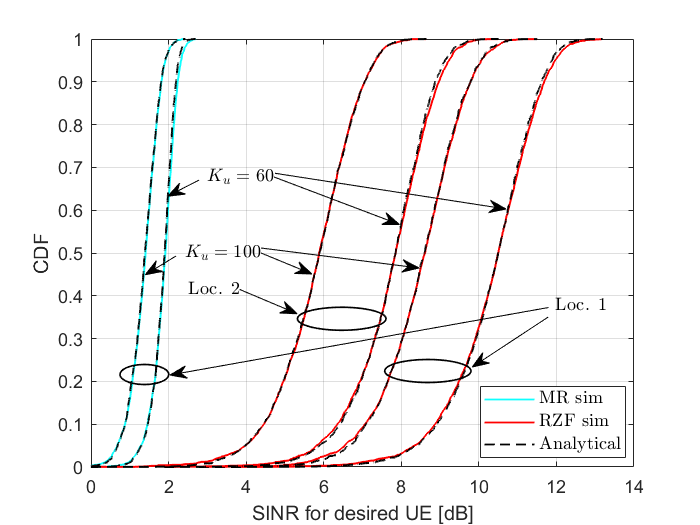}
\caption{Outage probability for different numbers of unknown interferers and different desired UE locations.}
\label{combined_SINR}
\vspace{-1em}
\end{figure}

Fig. \ref{combined_SINR} plots the CDF of the SINR at the desired UE with both the analytical model and Monte Carlo simulation results with different numbers of unknown interferers and different desired UE locations. The Fourier integral in \eqref{inv_formula} is approximated by a Riemann sum and computed efficiently using the Fast Fourier Transform (FFT) algorithm. We can see that the analytical performance tightly matches with the exact simulations for all simulated scenarios. We focus on RZF combining where the residual known interference is seen to be weaker than the unknown interference, which means that the unknown interference will be the dominant factor in the SINR's denominator. This results in a significant difference between the SINR curves when changing the number of unknown interferers for RZF, as well as larger variations within each curve caused by the unknown interference, which may potentially lead to outages. For the case of MR, the denominator of the SINR is dominated by the known interference from UEs belonging to the same serving AP cluster as the desired UE since it does not attempt to suppress the known interference. This results in lower SINRs in general with smaller SINR variations due to the unknown interference.

\begin{figure}
\centering
\setlength{\abovecaptionskip}{0.33cm plus 0pt minus 0pt}
\vspace{-0.35em}
\includegraphics[scale=0.47]{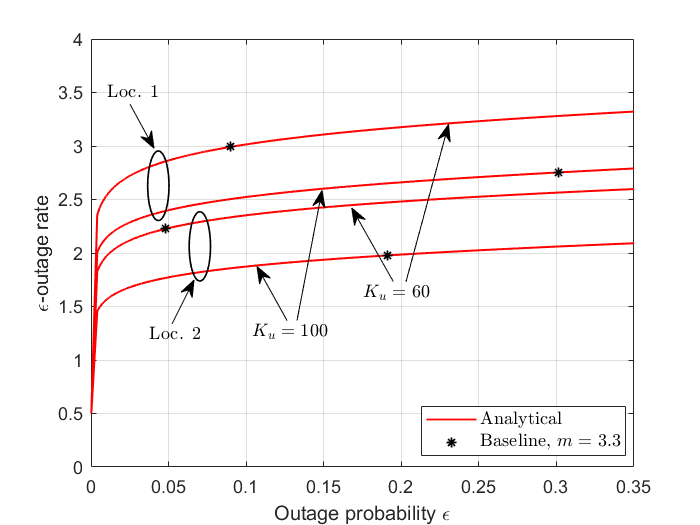}
\caption{$\epsilon$-outage SE with RZF.}
\label{combined_outage}
\vspace{-1em}
\end{figure}

In Fig. \ref{combined_outage}, we demonstrate how our analytical results can be used for robust rate adaptation. The figure plots the $\epsilon$-outage UL SE with RZF for different numbers of unknown interferers and different desired UE locations. We develop a baseline scheme for comparison, based on the classic idea of having a fixed fade margin. The baseline utilizes no knowledge of the unknown interference for rate adaptation. Alternatively, it divides the effective SINR of the desired UE in \eqref{SINR}, excluding the unknown interference term $\mathrm{IUI}_k^u$, by a fixed margin $m$ to make up for the effect of the unknown interference from UEs in neighboring clusters. The SE calculation for the baseline scheme is detailed in \cite[Eq.~(20)]{zaher2023bayesian}.

It is clear that the baseline scheme cannot effectively determine the UL transmission rate to maintain a specific target outage probability. This is due to the fact that changing the number of unknown interferers $K_u$ or the location of the desired UE results in a significant change in the outage probability. However, our proposed analytical model is able to determine the correct SE that results in any target predefined outage probability; thus, providing an effective tool for UL rate adaptation with guaranteed target outage performance.

\section{Conclusions}

This paper addresses the problem of outages caused by unknown UL interference in a cell-free massive MIMO network with LSFD. We derive an analytical model for the total unknown interference power distribution at the CPU and validate the accuracy of the analysis numerically. The model can be used for robust rate adaptation to handle outages caused by rapidly changing inter-cluster interference, which cannot be measured in practice. We have presented a baseline scheme for comparison, which utilizes a fixed margin to compensate for the unknown interference power. The numerical results show that with such a scheme, a minor change in the selected UL SE for transmission or a change in the simulated scenario leads to a significant change in the resulting outage probability. On the other hand, our framework is capable of always choosing the correct SE that guarantees a predetermined target outage. 

\section*{Appendix A}
The result follows from \cite[Th.~1]{zaher2023bayesian} that the unknown interference power at a single AP has an Inverse-Gamma distribution with probability density given by
\begin{equation}
    f_{\textrm{IUI}_{kl}^u}\left(x\right) = \frac{\beta_l^{\frac{3}{2}}}{\Gamma\left(\alpha_l\right)} \left(\frac{1}{x}\right)^{\alpha_l + 1}e^{-\frac{\beta_l}{x}}, \qquad x > 0.
\end{equation}

Utilizing the result presented in \cite{witkovsky2001computing} that
\begin{equation}
    \int_0^{\infty}x^{\alpha - 1}e^{-px-\frac{q}{x}}dx = 2\left(\frac{q}{p}\right)^{\alpha/2}K_{\alpha}\left(2\left(pq\right)^{1/2}\right).
\end{equation}

The characteristic function of $\textrm{IUI}_{kl}^u \sim \textrm{IG}\left(\alpha_l, \beta_l\right)$ can thus be computed as
\begin{equation}
    \phi_l\left(t\right) = \mathbb{E}\left\{e^{-jtX}\right\} = \frac{2\left(-j\beta_lt\right)^{\alpha_l/2}}{\Gamma\left(\alpha_l\right)}K_{\alpha_l}\left(2\left(-j\beta_lt\right)^{1/2}\right),
    \label{char_fun}
\end{equation}
where $X$ in \eqref{char_fun} denotes the $\textrm{IUI}_{kl}^u$ random variable for notational convenience. Now using our approximation for the covariance matrix of the unknown interference power as previously explained, we have 
\begin{equation}
    \textrm{IUI}_k^u \approx \sum_{l = 1}^{L_s}|a_{kl}|^2\textrm{IUI}_{kl}^u
\end{equation}
with $\{\textrm{IUI}_{kl}^u: l = 1, \hdots, L_s\}$ being independent. Using the properties of characteristic functions of sums of independent random variables, the characteristic function for the total unknown interference power can then be directly computed by \eqref{char_fun_tot}. Finally, the inversion theorem in \cite{gil1951note} is used to compute the CDF in \eqref{inv_formula} which concludes the proof.

\section*{References}
\renewcommand{\refname}{ \vspace{-\baselineskip}\vspace{-1.1mm} }
\bibliographystyle{ieeetr}
\bibliography{papercites}

\end{document}